\documentclass[a4paper]{jpconf}
\usepackage{graphicx}
\begin{document}
\title{Investigating the 2$^{nd}$ knee: The KASCADE-Grande experiment}

\author{
A.~Haungs$^\ast$, 
W.D.~Apel$^\ast$, 
F.~Badea$^{\ast,}$\footnote[1]{on leave of absence from National Institute of 
Physics and Nuclear Engineering, 7690 Bucharest, Romania},
K.~Bekk$^\ast$, 
A.~Bercuci$^\dag$,
M.~Bertaina$^\ddag$, 
J.~Bl\"umer$^{\ast,\S}$,
H.~Bozdog$^\ast$,
I.M.~Brancus$^\dag$,
M.~Br\"uggemann$^\P$,
P.~Buchholz$^\P$,
A.~Chiavassa$^\ddag$,
K.~Daumiller$^\ast$, 
F.~Di~Pierro$^\ddag$,
P.~Doll$^\ast$, 
R.~Engel$^\ast$,
J.~Engler$^\ast$, 
P.L.~Ghia$^\|$,
H.J.~Gils$^\ast$,
R.~Glasstetter$^{\ast\ast}$, 
C.~Grupen$^\P$,
D.~Heck$^\ast$, 
J.R.~H\"orandel$^\S$, 
K.-H.~Kampert$^{\ast\ast}$,
H.O.~Klages$^\ast$, 
Y.~Kolotaev$^\P$,
G.~Maier$^{\ast,}$\footnote{now at University Leeds, LS2~9JT~Leeds, United
Kingdom},
H.J.~Mathes$^\ast$, 
H.J.~Mayer$^\ast$, 
J.~Milke$^\ast$, 
B.~Mitrica$^\dag$,
C.~Morello$^\|$,
M.~M\"uller$^\ast$, 
G.~Navarra$^\ddag$,
R.~Obenland$^\ast$,
J.~Oehlschl\"ager$^\ast$, 
S.~Ostapchenko$^{\ast,}$\footnote{on leave of absence from Moscow State University, 
119899~Moscow, Russia}, 
S.~Over$^\P$,
M.~Petcu$^\dag$, 
T.~Pierog$^\ast$, 
S.~Plewnia$^\ast$,
H.~Rebel$^\ast$, 
A.~Risse$^{\dag\dag}$, 
M.~Roth$^\S$, 
H.~Schieler$^\ast$, 
O.~Sima$^\dag$, 
M.~St\"umpert$^\S$, 
G.~Toma$^\dag$, 
G.C.~Trinchero$^\|$,
H.~Ulrich$^\ast$,
S.~Valchierotti$^\ddag$,
J.~van~Buren$^\ast$,
W.~Walkowiak$^\P$,
A.~Weindl$^\ast$,
J.~Wochele$^\ast$, 
J.~Zabierowski$^{\dag\dag}$,
S.~Zagromski$^\ast$, and
D.~Zimmermann$^\P$}

\address{$^\ast$Institut\ f\"ur Kernphysik, Forschungszentrum Karlsruhe,
76021~Karlsruhe, Germany}
\address{$^\dag$National Institute of Physics and Nuclear Engineering,
7690~Bucharest, Romania}
\address{$^\ddag$Dipartimento di Fisica Generale dell'Universit{\`a},
10125 Torino, Italy}
\address{$^\S$Institut f\"ur Experimentelle Kernphysik,
Universit\"at Karlsruhe, 76021 Karlsruhe, Germany}
\address{$^\P$Fachbereich Physik, Universit\"at Siegen, 57068 Siegen, 
Germany}
\address{$^\|$Istituto di Fisica dello Spazio Interplanetario, CNR, 
10133 Torino, Italy}
\address{$^{\ast\ast}$Fachbereich Physik, Universit\"at Wuppertal, 42097
Wuppertal, Germany}
\address{$^{\dag\dag}$Soltan Institute for Nuclear Studies, 90950~Lodz, 
Poland}

\ead{haungs@ik.fzk.de}

\begin{abstract}
Recent results from the multi-detector set-up KASCADE on measurements 
of cosmic rays in the energy range of the so called "first" knee 
(at $\approx 3\,$PeV) indicate a distinct knee in the energy spectra 
of light primary cosmic rays and an increasing dominance of heavy 
ones towards higher energies. 
This leads to the expectation of knee-like features 
of the heavy primaries at around 100 PeV. To investigate this energy 
region KASCADE has recently been extended by a factor 10 in area to the 
new experiment KASCADE-Grande. Main results of KASCADE as well as 
set-up, capabilities, and status of KASCADE-Grande are presented. 
\end{abstract}

\section{Introduction}

\begin{figure*}[t]
\centering
\vspace*{0.1cm}
\includegraphics[width=125mm]{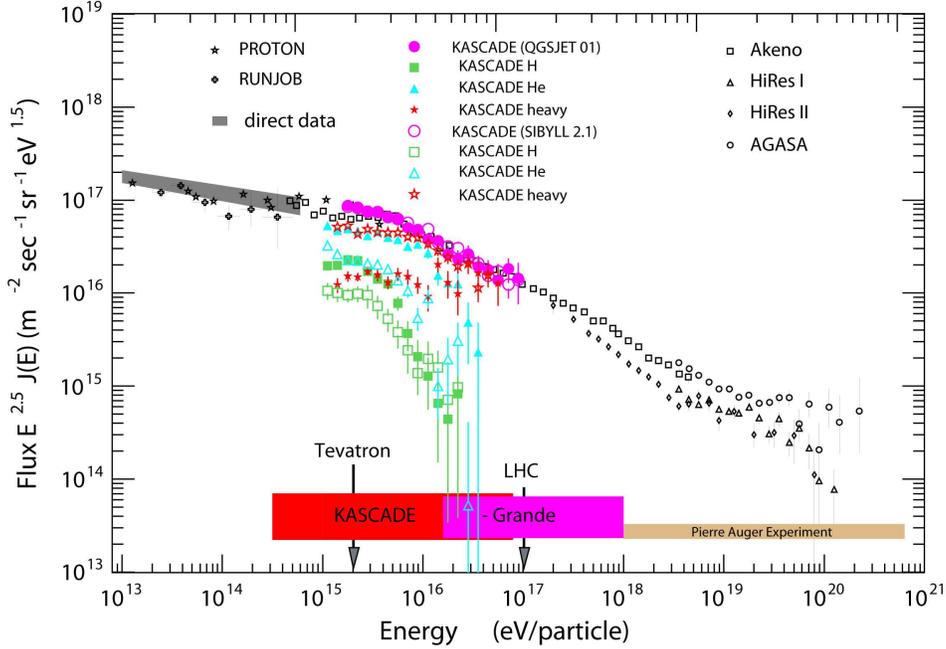}
\caption{Primary cosmic ray flux and primary energy range covered by 
KASCADE-Grande. Results of the KASCADE data analyses are also shown 
(see text).}
\label{knee}
\end{figure*}

The all-particle energy spectrum of cosmic rays shows a distinctive
discontinuity at few PeV, known as the knee, where the spectral index
changes from $-2.7$ to approximately $-3.1$~(Fig.~\ref{knee}). 
At that energy direct
measurements are presently hardly possible due to the low flux, but 
indirect measurements observing extensive air showers (EAS) are 
performed. 
Astrophysical scenarios like the change of the acceleration 
mechanisms at the cosmic ray sources 
(supernova remnants, pulsars, etc.) 
or effects of the transport mechanisms inside the Galaxy (diffusion 
with escape probabilities) are conceivable for the origin of the knee
as well as particle physics reasons like a new kind of hadronic 
interaction inside the atmosphere or during the transport through 
the interstellar medium. Two
classes of theories (diffusion or acceleration based) predict 
knee positions occurring at constant rigidity of the particles. 
On the other hand, the hypothesis of new hadronic
interaction mechanisms at the knee energy, as for example the 
production of heavy particles in $pp$ collisions, implies an 
atomic mass 
dependence of the knee positions. 
It is obvious that only detailed measurements over the whole energy
range of the knee from $10^{14}\,$eV to $10^{18}\,$eV
and analyses of the primary energy spectra for the different incoming 
particle types can validate or disprove some of 
these models. 

The highest cosmic energies above the so called ankle 
are believed to be of extragalactic origin. 
Thus, in the experimentally less explored region between 
the first ("proton") knee and the ankle there are two more 
peculiarities of the cosmic ray spectrum expected: 
(i) A knee of the heavy component which is 
expected (depending on the model) either at the position of the 
first knee times Z (the charge) or times A (the mass) of iron. 
(ii) A transition region from galactic to extragalactic origin 
of cosmic rays, where there is no theoretical reason for a 
smooth crossover in slope and flux. \\ 

Despite EAS measurements with many experimental setups in the 
last five decades the origin of 
the knee is still not clear, as the disentanglement of the threefold 
problem of estimate of energy and mass plus the understanding of
the air-shower development in the Earth's atmosphere remains an 
experimental challenge. For a detailed discussion of the subject
see the review in~\cite{rpp}. 

The multi-detector system KASCADE-Grande (KArlsruhe Shower Core and Array 
DEtector and Grande array) approaches this challenge by 
measuring as much as possible redundant information from
each single air-shower event. 
The multi-detector arrangement allows 
to measure the total electron and muon numbers of the shower 
separately using an array of shielded and unshielded detectors. 
Additionally muon densities at further
three muon energy thresholds 
and the hadronic core of the 
shower by an iron sampling calorimeter are measured. 
Recently, the original KASCADE experiment~\cite{kas} was extended in 
area by a factor 10 to the new
experiment KASCADE-Grande~\cite{Navar04,Haung03,Haungs-KG04}.
KASCADE-Grande allows now a full coverage of the energy range 
around the knee, including the possible second 
knee (see Fig.~\ref{knee}). \\

From KASCADE~\cite{kas} measurements we do know that at a few 
times \mbox{10$^{15}$ eV} the knee is due to light 
elements~\cite{Anton02}, 
that the knee positions depend on the kind of the incoming particle, 
and that cosmic rays around the knee arrive our Earth 
isotropically~\cite{Gmaier1,Gmaier2}. 
KASCADE-Grande~\cite{Navar04,Haung03}, measuring higher energies, 
will prove, if existent, the knee corresponding to heavy elements.
Additionally KASCADE could show that no current
hadronic interaction model which are unavoidably needed for the 
interpretation of air shower data, describes very well 
cosmic ray measurements in the energy range of the knee and 
above~\cite{Ulric05,Ulric04}. 
These model uncertainties are
due to the lack of accelerator data at these energies and especially 
for the forward direction of collisions. Multi-detector systems like 
KASCADE and KASCADE-Grande offer the possibility
of testing and tuning the different hadronic interaction models.

With its capabilities KASCADE-Grande is also the ideal testbed for 
the development and calibration of new air-shower
detection techniques like the measurement of EAS radio
emission~\cite{lopes}.

The present contribution will summarize the main results of the KASCADE
experiment and discuss the capabilities and status of KASCADE-Grande.

\section{The KASCADE-Grande Experiment}

The KASCADE-Grande experiment, located at the 
Forschungszentrum Karlsruhe, Germany, 
(49$^\circ$n, 8$^\circ$e, 110$\,$m$\,$a.s.l.)~measures 
showers in a primary energy range from $100\,$TeV to $1\,$EeV 
and provides 
multi-parameter measurements on a large number of observables 
concerning electrons, muons at 4 energy thresholds, and hadrons.
The main detector components of KASCADE-Grande are the KASCADE Array, 
the Grande array, the Central Detector, and the 
Muon Tracking Detector (see Table~\ref{tab1}). 

The KASCADE Array measures the 
total electron and muon numbers ($E_\mu>230\,$MeV)
of the shower separately using an array of 252 detector stations 
containing shielded and unshielded detectors at the same place 
in a grid of $200 \times 200\,$m$^2$.
The excellent time resolution of these detectors allows also decent 
investigations of the arrival directions of the showers in searching
large scale anisotropies and, if existent, cosmic ray point sources. 
The KASCADE array is optimized to measure EAS in the energy range of 
100 TeV to 80 PeV.

The Muon Tracking Detector ($128\,$m$^2$) measures the incidence 
angles of muons ($E_\mu > 800\,$MeV) relative to the shower arrival 
direction. These measurements provide a sensitivity to the
longitudinal development of the showers. 

\begin{figure}[h]
\begin{minipage}{18pc}
\centering
\includegraphics[width=17pc]{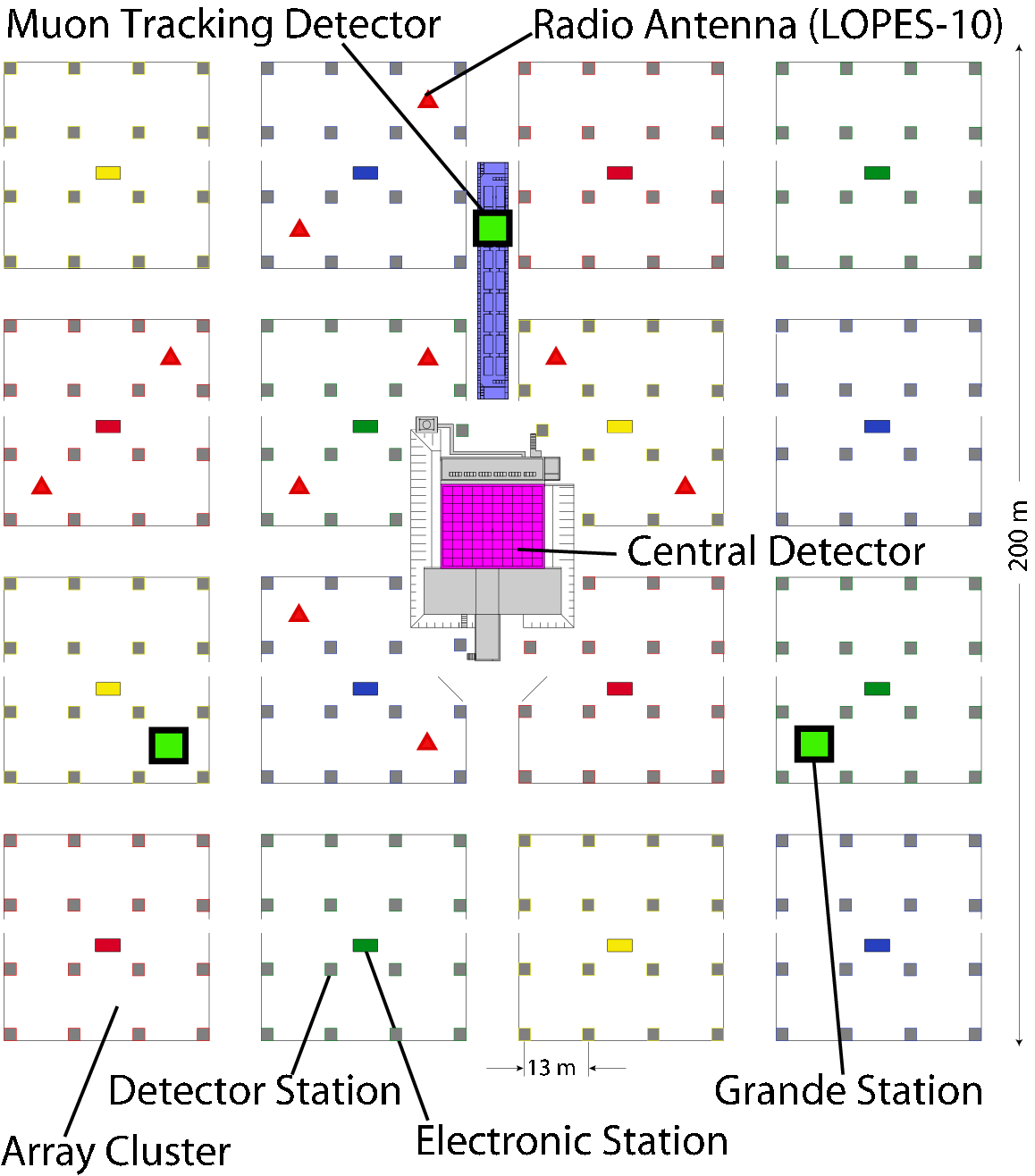}
\caption{\label{KASCADE_radio}
The main detector components of the KASCADE experiment: 
(the 16 clusters of) Field Array, Muon Tracking Detector and 
Central Detector. The location of 10 radio antennas is also 
displayed, as well as three stations of the Grande array.}
\end{minipage}\hspace{2pc}%
\begin{minipage}{18pc}
\includegraphics[width=18pc]{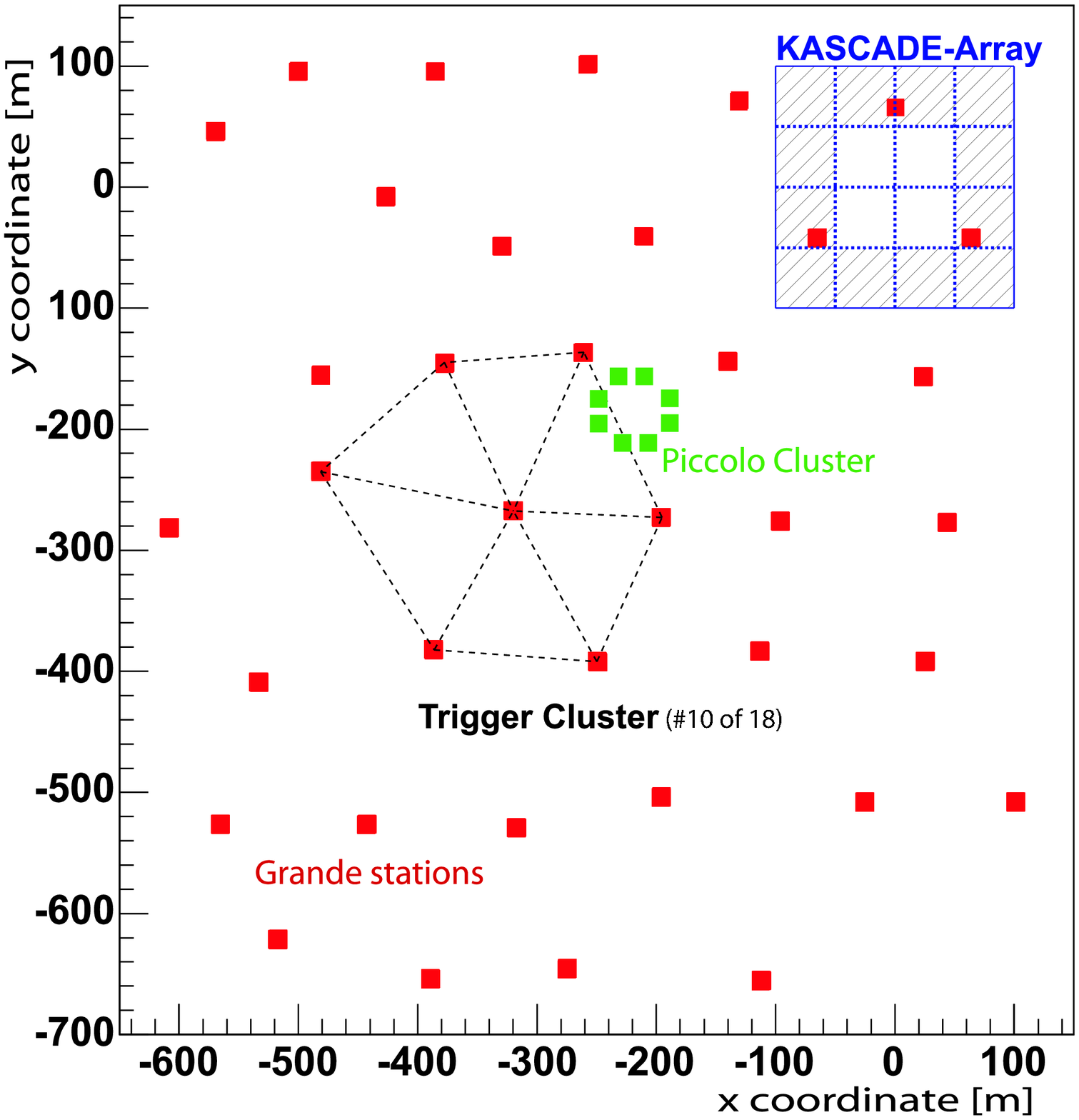}
\caption{\label{grande}Layout of the KASCADE-Grande experiment
with its 37 Grande and 8 Piccolo stations. One of the 18 Grande
trigger hexagons is also shown.}
\end{minipage} 
\end{figure}
\begin{table}[h]
\caption{\label{tab1}Compilation of the main KASCADE-Grande 
detector components. The threshold values are given for the 
particles kinetic energy.}
\begin{center}
\begin{tabular}{l|l|c}
\hline
Detector                & Particles                            
&  sensitive area [m$^2$]        \\
\hline
Grande                  & charged             
& 370  \\
Piccolo     		& charged            
& 80  \\
KASCADE array $e/\gamma$& electrons                
& 490  \\
KASCADE array $\mu$	& muons ($E_\mu^{\rm thresh}=230\,$MeV) 
& 622  \\
MTD                     & muons ($E_\mu^{\rm thresh}=800\,$MeV) 
& 3$\times$128   \\
Trigger Plane		& muons ($E_\mu^{\rm thresh}=490\,$MeV) 
& 208  \\
MWPCs/LSTs		& muons ($E_\mu^{\rm thresh}=2.4\,$GeV) 
& 3$\times$129 \\
Calorimeter  		& hadrons ($E_h^{\rm thresh}=10-20\,$GeV)
& 9$\times$304  \\
LOPES	  		& radio wave ($40-80\,$MHz)
& 30 dipole antennas \\
\hline
\end{tabular}
\end{center}
\end{table}
The hadronic core of the shower is measured by a $300\,$m$^2$ 
iron sampling calorimeter installed at the KASCADE Central Detector: 
Three other components - trigger plane (serves also as timing
facility), multiwire proportional chambers (MWPC), 
and limited streamer tubes (LST) - offer additional 
valuable information on the penetrating muonic component 
at $490\,$MeV and $2.4\,$GeV energy thresholds, respectively. 

The multi-detector concept of the KASCADE experiment
(Fig.~\ref{KASCADE_radio}, which is 
operating since 1996 has been translated to higher primary 
energies through KASCADE-Grande~\cite{KG-cris}.

The 37 stations of the Grande Array (Fig.~\ref{grande})  
extend the cosmic ray measurements up to primary 
energies of \mbox{1 EeV}. 
The Grande stations, \mbox{10 m$^2$} of plastic scintillator 
detectors each, are spaced at approximative \mbox{130 m} 
covering a total area of \mbox{$\sim$ 0.5 km$^2$}. 
There are 16 scintillator sheets in a station read out by 16 
high gain photomultipliers; 4 of the scintillators are read out 
also by 4 low gain PMs. The covered dynamic range is 
up to \mbox{3000 mips/m$^2$}. A trigger signal is build when 
7 stations in a hexagon (trigger cluster, see 
Fig.~\ref{grande}) are fired. Therefore, the Grande array 
consists of 18 hexagons with a total trigger rate of \mbox{0.5 Hz}.

Additionally to the Grande Array a compact array, named Piccolo, has 
been build in order to provide a fast trigger to KASCADE ensuring 
joint measurements for showers with cores located far from the 
KASCADE array.
The Piccolo array consists of 8 stations with \mbox{11 m$^2$} plastic 
scintillator each, distributed over an area of \mbox{360 m$^2$}. 
One station contains 12 plastic scintillators organized in 6 modules; 
3 modules form a so-called electronic station providing 
ADC and TDC signals. 
A Piccolo trigger is built and sent to KASCADE and Grande
when at least 7 out of the 48 modules of Piccolo are fired. 
Such a logical condition leads to a trigger rate of \mbox{0.3 Hz}.

To improve further the data quality a self-triggering, 
dead-time free FADC-based DAQ system will be implemented in 
order to record the full time evolution of energy deposits in 
the Grande stations at an effective sampling rate of 
\mbox{250 MHz} and high resolution of 12 bits in two gain 
ranges~\cite{Andre04}. This will lead to an intrinsic 
electron-muon separation of the data signal at the Grande array. 

The whole KASCADE-Grande setup is read out if a certain multiplicity
of the KASCADE, Piccolo, or the Grande array detector stations 
or of the trigger plane is firing, leading to a total trigger rate 
of $\approx 4.5\,$Hz.

The redundant information of the showers measured by the 
Central Detector and the Muon Tracking Detector
is predominantly being used for tests and improvements 
of the hadronic interaction models underlying the 
analyses~\cite{isv04}.  

For the calibration of the radio signal emitted by the air shower in
the atmosphere an array of first 10 and meanwhile 30 dipole antennas 
(LOPES) is set up on the site of the 
KASCADE-Grande experiment~\cite{lopes-texas}.

\section{KASCADE results}

\subsection{Search for anisotropies and point sources}

\begin{figure}[h]
\begin{minipage}{18pc}
\centering
\includegraphics[width=17pc]{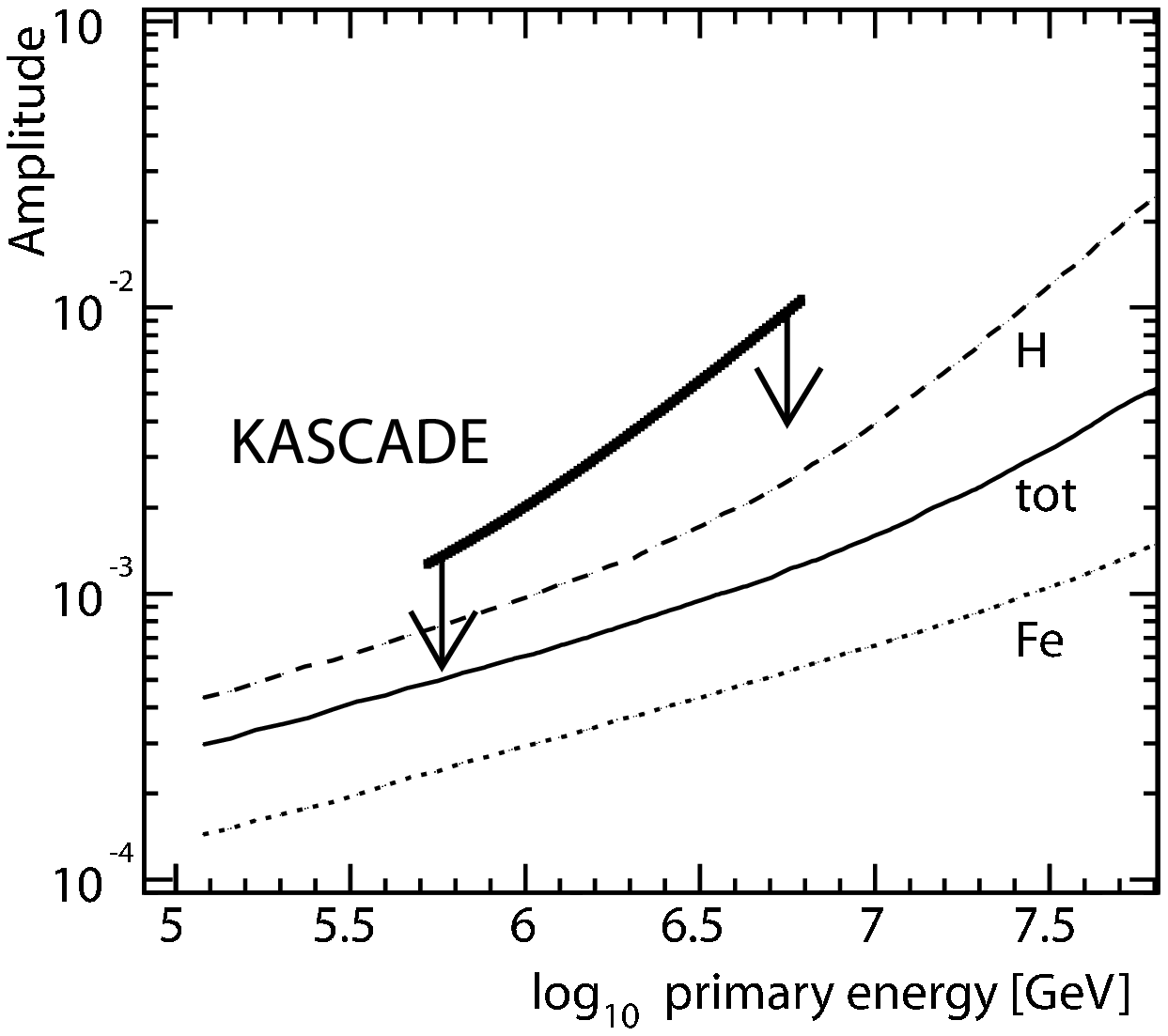}
\caption{\label{an-ls}Rayleigh amplitude of the harmonic analyses 
of the KASCADE data~\cite{Gmaier1} (limit on a 95\% 
confidence level) compared to theory predictions~\cite{candia}.
}
\end{minipage}\hspace{2pc}%
\begin{minipage}{17pc}
\includegraphics[width=17pc]{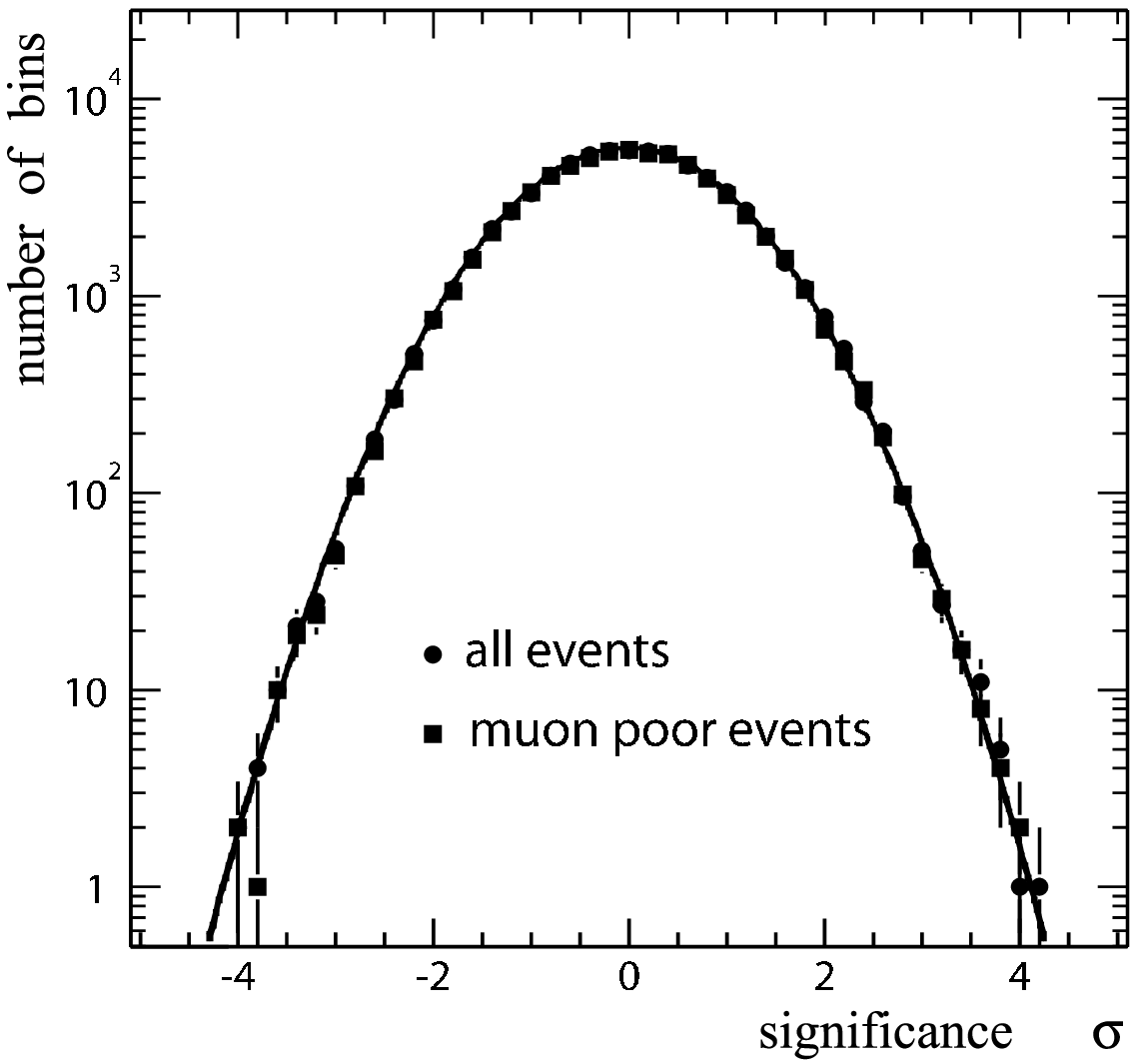}
\caption{\label{an-ps}Significance distributions for searching
point sources on the sky map seen by the KASCADE 
experiment~\cite{Gmaier2}.}
\end{minipage} 
\end{figure}

Investigations of anisotropies in the arrival directions of the 
cosmic rays give additional information on the cosmic ray origin 
and of their propagation.  
Depending on the model of the origin of the knee and on 
the assumed structure of the galactic magnetic field
one expects large-scale anisotropies on a scale of 
$10^{-4}$ to $10^{-2}$ in the energy region of the knee.
The limits of large-scale anisotropy analyzing the KASCADE data as
shown in Fig.~\ref{an-ls} 
are determined to be between $10^{-3}$ at 0.7 PeV primary energy 
and $10^{-2}$ at 6 PeV~\cite{Gmaier1}. 
These limits were obtained by investigations of 
the Rayleigh amplitudes and phases of the first harmonics. 
Taking into account possible nearby sources of galactic cosmic 
rays like the Vela Supernova 
remnant~\cite{ptuskin} the limits of KASCADE already exclude 
particular model predictions. 

The interest for looking to point sources in the KASCADE data sample 
arises from the possibility of unknown near-by sources, where the 
deflection of the charged cosmic rays would be small or by sources 
emitting neutral particles like high-energy gammas or neutrons. 
The scenario for neutrons is very interesting for KASCADE-Grande,
since the neutron decay length at these energies is in the order of
the distance to the Galactic center. 
In KASCADE case, the full sample of air showers were investigated 
as well as a sample of "muon-poor" showers which is a 
sample with an enhanced number  of candidates of $\gamma$-ray
induced events (Fig.~\ref{an-ps}). 
No significant excess was found in both
samples~\cite{Gmaier2}.  

\subsection{Energy spectra of individual mass groups}

\begin{figure}[t]
\centering
\vspace*{0.1cm}
\includegraphics[width=100mm]{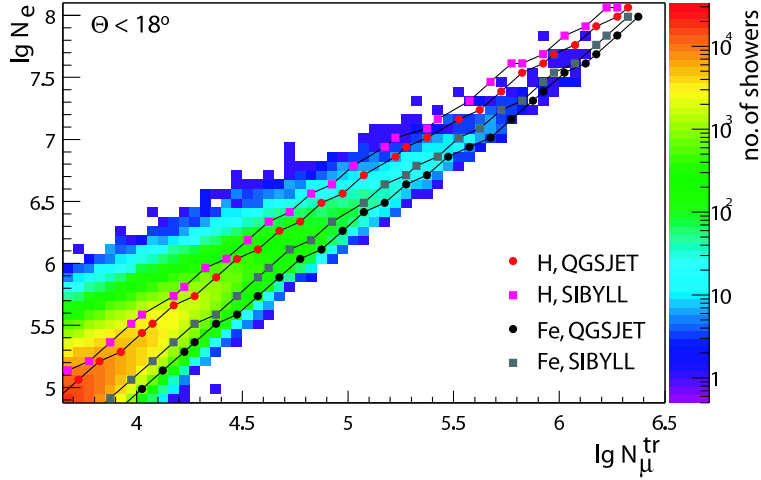}
\caption{Two dimensional electron ($N_e$)~vs.~muon 
($N_\mu^{\rm tr}\,=\,$number of muons in 40-200m core distance) 
number spectrum measured by the KASCADE array. 
The lines display the most probable values for proton
and iron primaries obtained by CORSIKA simulations employing 
different hadronic interaction models.} 
\label{data}
\end{figure}

The KASCADE data analyses aims to reconstruct the energy spectra 
of individual mass groups taking into account not only different
shower observables, but also their correlation on an event-by-event
basis.
The content of each cell of the two-dimensional spectrum of 
reconstructed electron number~vs.~muon number (Fig.~\ref{data}) 
is the sum of contributions from the individual primary elements.
Hence the inverse problem 
{\small $g(y) = \int{K(y,x)p(x)dx}$} 
with {\small $y=(N_e,N_\mu^{\rm tr})$} and {\small $x=(E,A)$} 
has to be solved.
This problem results
in a system of coupled Fredholm integral equations of the form \\
{\small  $\frac{dJ}{d\,\lg N_e\,\,d\,\lg N_\mu^{tr}} = 
 \sum_A \int\limits_{-\infty}^{+\infty} \frac{d\,J_A}{d\,\lg E} 
  \cdot 
  p_A(\lg N_e\, , \,\lg N_\mu^{tr}\, \mid \, \lg E)
  \cdot 
  d\, \lg E $ } \\
where the probability $p_A$  
is a further integral with the kernel function 
{\small $k_A = r_A \cdot \epsilon_A \cdot s_A$}
factorized into three parts. The quantity $r_A$ describes 
the shower fluctuations, 
i.e. the distribution of electron and muon number for given 
primary energy and mass. The quantity $\epsilon_A$ describes 
the trigger efficiency of the experiment, 
and $s_A$ describes the reconstruction
probabilities, i.e. the distribution of reconstructed $N_e$ and 
$N_\mu^{\rm tr}$ for given true numbers of electrons and muons.
The probabilities $p_A$ are obtained by 
Monte Carlo simulations on basis of 
two different hadronic interaction models (QGSJET$\,01$~\cite{qgs}, 
SIBYLL$\,2.1$~\cite{sib}) as options embedded in 
CORSIKA~\cite{cors}. 
\begin{figure}[h]
\begin{minipage}{18pc}
\includegraphics[width=18pc]{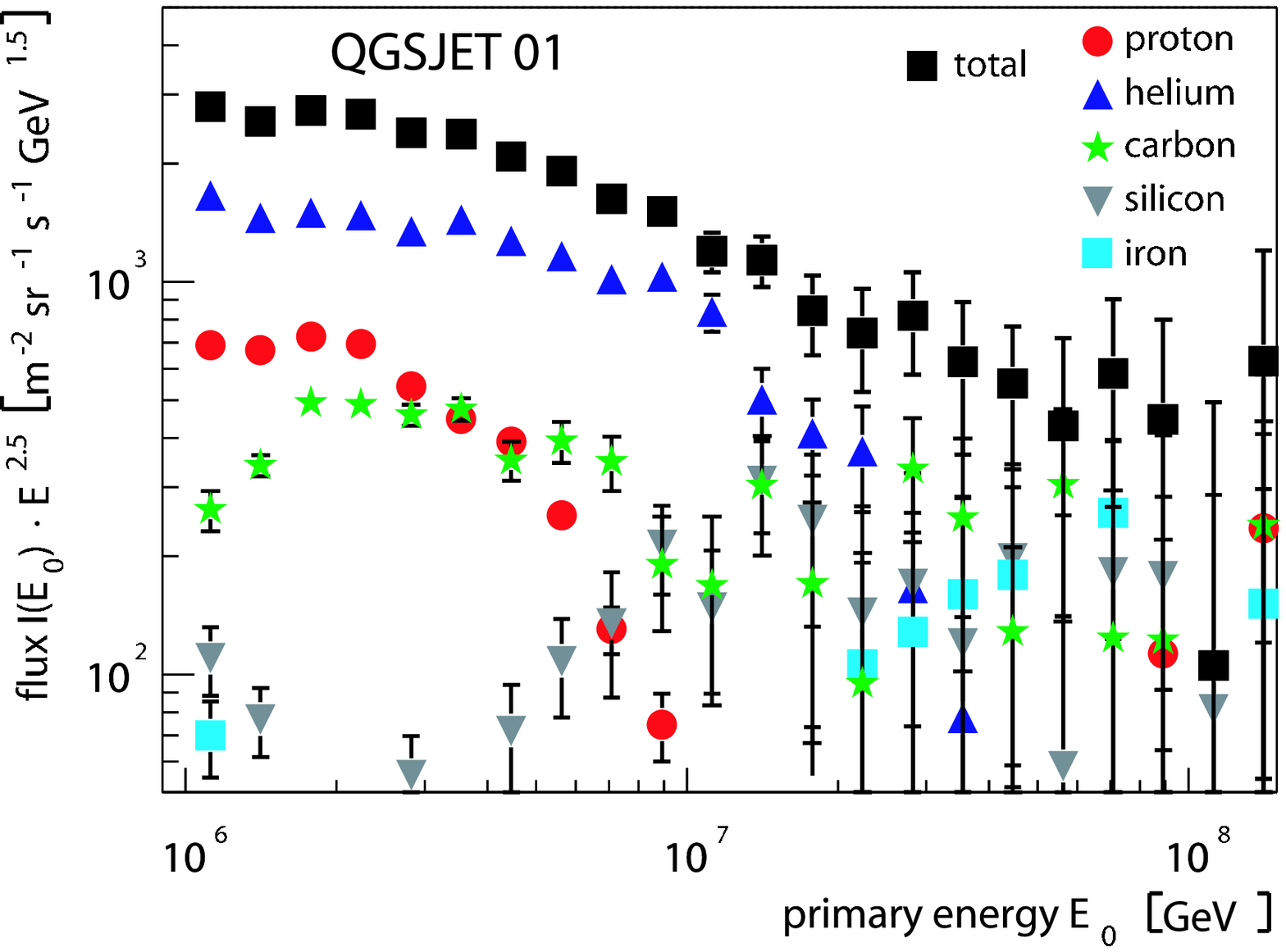}
\caption{\label{spe-qgs}Result of the unfolding procedure based
on QGSJET$\,01$.}
\end{minipage}\hspace{2pc}%
\begin{minipage}{18pc}
\includegraphics[width=18pc]{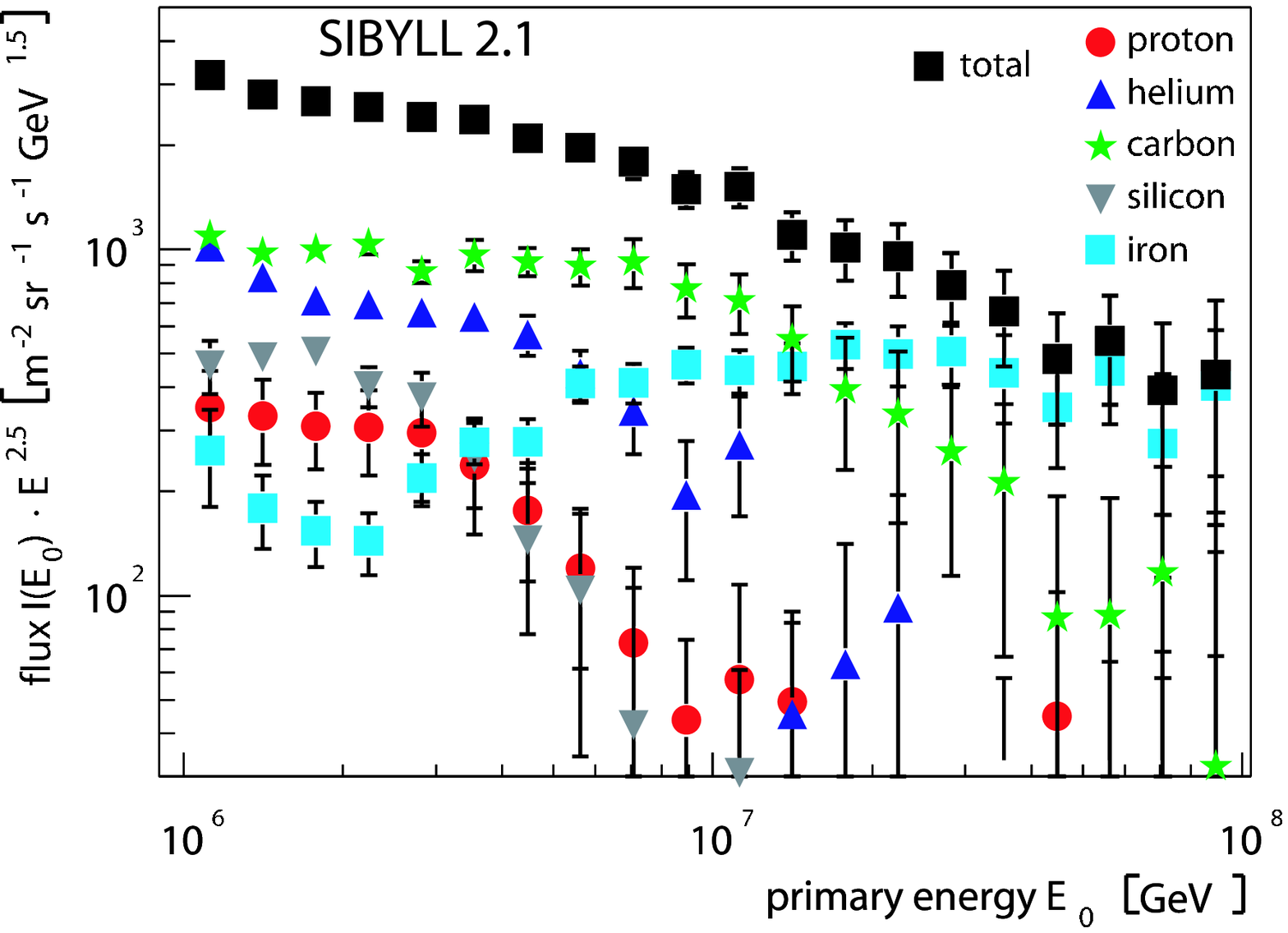}
\caption{\label{spe-sib}Result of the unfolding procedure based on SIBYLL$\,2.1$.}
\end{minipage} 
\end{figure}
By applying the above described procedures (with the assumption of
five primary mass groups, only) to the experimental 
data energy spectra are obtained as displayed in 
Figs.~\ref{spe-qgs},~\ref{spe-sib} and in Fig.~\ref{knee}, where 
the resulting spectra for primary oxygen, silicon,
and iron are summed up for a better visibility. 

A knee like feature is clearly visible in the all particle spectrum,
which is the sum of the unfolded single mass group spectra, 
as well as in the spectra of primary proton and helium.
This demonstrates that the elemental composition of cosmic
rays is dominated by the light components below the knee and 
by a heavy component above the knee feature. Thus, the knee feature 
originates from a decreasing flux of the light primary 
particles~\cite{Ulric05,Ulric04}.

\subsection{Inaccuracies of hadronic interaction models}

Comparing the unfolding results based on the two different hadronic 
interaction models, the model dependence when interpreting the data
is obvious. 
Modeling the hadronic 
interactions underlies assumptions from particle physics theory and 
extrapolations resulting in large uncertainties, which are reflected 
by the discrepancies of the results presented here. 
In Fig.~\ref{data} the predictions of the $N_e$ and $N_\mu^{tr}$ 
correlation for the 
two models are overlayed to the measured distribution
in case of proton and iron primaries. 
It is remarkable that all four lines have a more or less parallel 
slope which is different from the data distribution. There, the knee is 
visible as kink to a flatter $N_e$-$N_\mu^{tr}$ 
dependence above $\lg N_\mu^{tr} \approx 4.2\,$. The heavier primary contribution 
on the results based on the SIBYLL model is due to predictions of
a smaller ratio of muon to electron number for all primaries. 
Comparing the residuals of the unfolded two dimensional distributions for 
the different models with the initial data set 
we conclude~\cite{Ulric04} 
that at lower energies the SIBYLL model and at higher energies 
the QGSJET model are able to describe 
the correlation consistently, but none of the present models
gives a contenting description of the whole data set. 
These findings are confirmed by detailed investigations of further
shower observables measured by KASCADE~\cite{isv04}.

Crucial parameters in the modeling of hadronic 
interaction models which can be responsible for these inconsistencies 
are the total nucleus-air cross-section and the parts of the inelastic and 
diffractive cross sections leading to shifts of the position of the 
shower maximum in the atmosphere and, therefore, to a change of the muon
and electron numbers as well as to their correlation on single air 
shower basis. 
The multiplicity of the pion generation at all energies at 
the hadronic interactions during the air shower development is also
a 'semi-free' parameter in the air-shower modeling as accelerator data
have still large uncertainties, in particular for the forward
direction~\cite{engel}.

\section{Status, capabilities and perspectives of KASCADE-Grande}

\begin{figure}[hb]
\begin{minipage}{18pc}
\includegraphics[width=18pc]{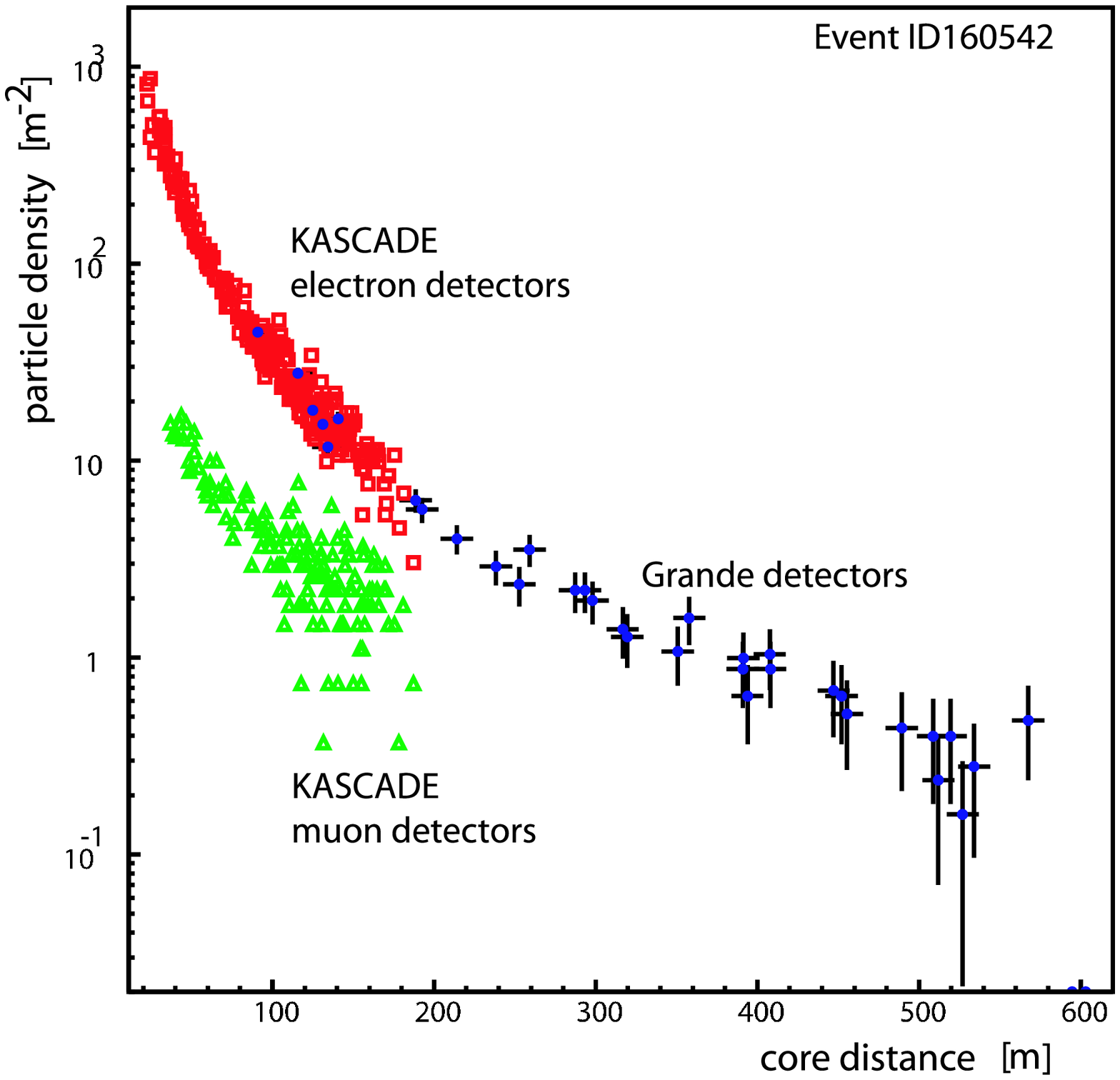}
\caption{\label{LDFexample}Particle densities in the different detector types 
of KASCADE-Grande measured for a single event.}
\end{minipage}\hspace{2pc}%
\begin{minipage}{18pc}
\centering
\includegraphics[width=17pc]{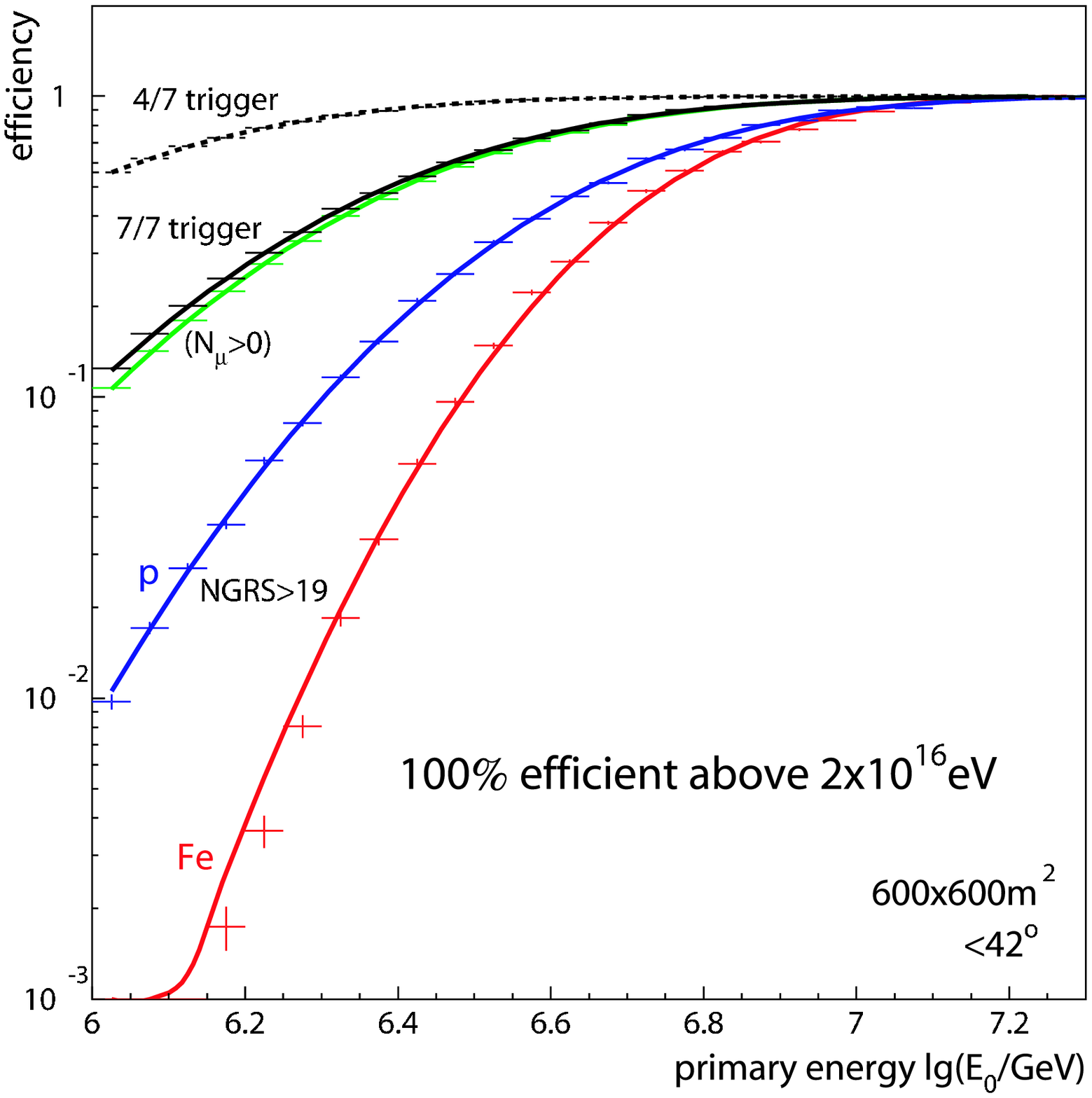}
\caption{\label{effi}Efficiency of the Grande array 
(details see text). CORSIKA simulations including detailed simulation
of the detector response.}
\end{minipage} 
\end{figure}

\begin{figure}[ht]
\begin{minipage}{20pc}
\includegraphics[width=20pc]{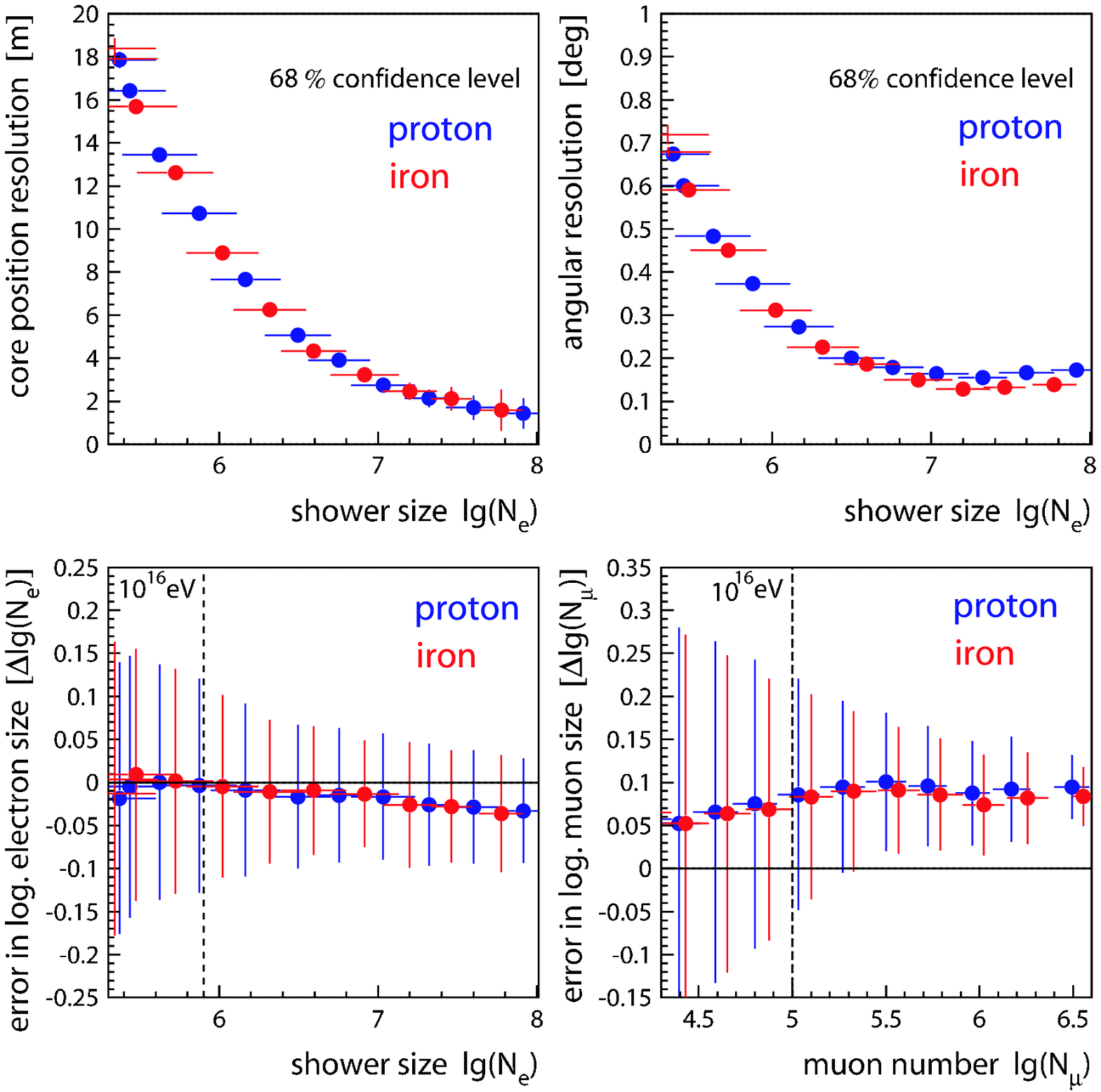}
\caption{Core, angular, and shower size reconstruction resolution
of the KASCADE-Grande experiment. CORSIKA simulations including 
detailed simulation of the detector response.}
\label{KG-res}
\end{minipage}\hspace{2pc}%
\begin{minipage}{16pc}
\includegraphics[width=16pc]{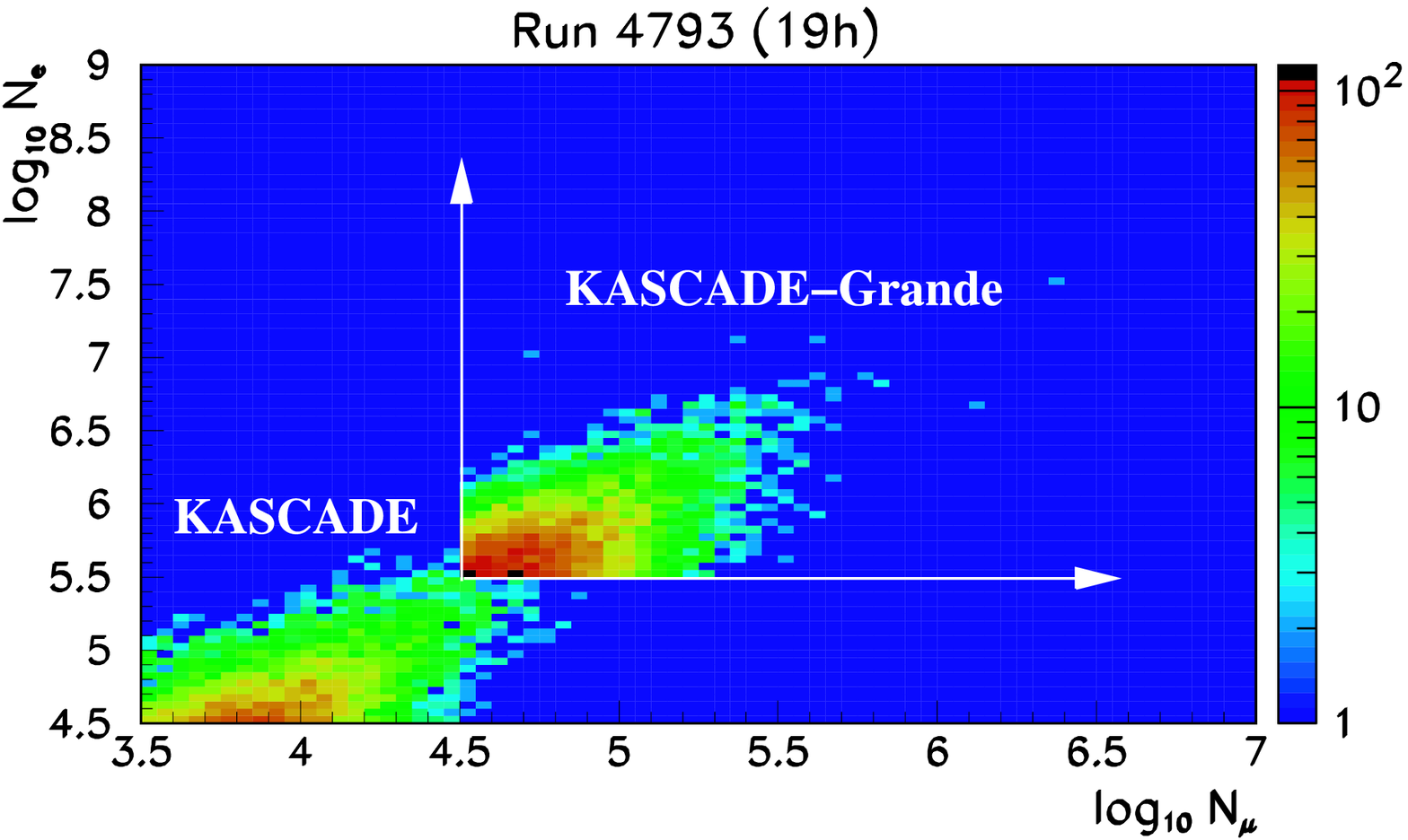}
\caption{\label{april04_data}Comparison between KASCADE 
and KASCADE-Grande data for a combined test-run.}
\end{minipage} 
\end{figure}

Fig.~\ref{LDFexample} shows, for a single event,
the lateral distribution of electrons and muons 
reconstructed with KASCADE and the charge particle densities 
measured by the Grande stations. This example illustrates the 
capabilities of KASCADE-Grande and the high quality of the data. 
The KASCADE-Grande reconstruction procedure follows iterative 
steps: shower core position, 
angle-of-incidence, and total number of charged particles are 
estimated from Grande Array data; 
the muon densities and with that the reconstruction of the
total muon number is provided by the KASCADE muon detectors.
The reconstruction accuracy (Fig.~\ref{KG-res}) of the
shower core position and direction is in the order of 
\mbox{4 m} (\mbox{13 m})
and $0.18^{\circ}$ ($0.32^{\circ}$) with $68$\% ($95$\%) 
confidence level for
proton and iron showers at \mbox{100 PeV} primary energy and 
$22^{\circ}$ zenith angle~\cite{Glass03}. 
The statistical uncertainty of the
shower sizes are around $10-15$\% and $20-25$\% for the 
total numbers of electrons and muons, respectively. 
The critical point of the KASCADE-Grande reconstruction 
is the estimation of the muon number due to the 
limited sampling of the muon lateral 
distribution by the KASCADE muon detectors.
The systematic uncertainty for the muon number depends on the 
radial range of the data measured by the KASCADE array and 
the chosen lateral distribution function. 

At the KASCADE experiment, the two-dimensional distribution 
shower size - truncated number of muons played the fundamental 
role in reconstruction of energy spectra of single mass groups. 
In Figure~\ref{april04_data} the correlation of these two shower 
sizes for both cases KASCADE and KASCADE-Grande measurements
are compared for a 1-day test-run. 
For the same run time, due to its 10 times larger area compared 
with KASCADE, the Grande Array sees a significant number 
of showers at primary energies \mbox{$\sim$10 times} higher. 
Hence, Figure~\ref{april04_data} illustrates the capability of
KASCADE-Grande to perform an unfolding procedure like in 
KASCADE. \\

Figure~\ref{effi} shows the efficiency characteristics of the 
KASCADE-Grande array. For internal tests of the detector stations 
a 4/7 trigger is performed at the hexagons. 
The efficiency of the 7/7 trigger is also shown which is
only marginal smaller if the additional requirement is imposed 
that the muon number has to be reconstructed with the 
information of the muon detectors of the original KASCADE array. 
To reduce efficiently the amount of 
data a software cut will be applied with the requirement of at 
least 20 Grande stations (NGRS$>$19) have to 
be fired. A hundred percent efficiency is than reached for all 
primary particle types for energies above $2 \cdot 10^{16}\,$eV, 
providing still a large overlap with the KASCADE energy range.
The limit at high energies for Grande is due to the limitation 
in area and not saturation of the detectors, as even at primary 
energy of $10^{18}\,$eV only one station in average is saturated.

KASCADE-Grande started end of 2003 with combined measurements
of all detector components. 
Currently (spring 2005) a FADC system is installed at the 
Grande stations which
will run in parallel to the present data acquisition. Besides the
physics gain with the possible intrinsic separation of electrons 
and muons by having the full shower time history, this additional
information will be used for cross-checking the calibration
procedures of KASCADE-Grande. 

\section{Conclusions}

The~extension~of~KASCADE~to~the~KASCADE-Grande experiment, 
accessing higher primary energies, 
is expected to solve the question of the existence 
of a knee-like structure  
corresponding to heavy elements. 
KASCADE-Grande keeps the multi-detector 
concept for tuning different interaction models at 
primary energies up to \mbox{$10^{18}$ eV}. 
KASCADE-Grande also provides the perfect environment 
detecting radio emission in extensive air showers, which 
is the aim of the LOPES project~\cite{lopes-nature}.

\medskip 
\noindent {\bf Acknowledgments} \\
KASCADE-Grande is supported by 
the Ministry for Research and Education of Germany,
the INFN of Italy,  
the Polish State Committee for Scientific Research 
(KBN grant for 2004-06) and the Romanian National Academy 
for Science, Research and Technology.

\smallskip

\end{document}